\documentclass[conference,UTF8]{IEEEtran}
\IEEEoverridecommandlockouts

\usepackage{amsfonts,amsmath,amssymb,verbatim,graphicx,bm,float,multirow,makecell,cite,url}
\usepackage{ulem,booktabs}
\usepackage[numbers, sort]{natbib}
\begin{document}

\title{Attentive Q-Matrix Learning for Knowledge Tracing\\
}

\author{\IEEEauthorblockN{1\textsuperscript{st} Zhongfeng Jia}
\IEEEauthorblockA{\textit{Lanzhou University}\\
Lanzhou, China \\
jiazhf21@lzu.edu.cn}
\and
\IEEEauthorblockN{2\textsuperscript{nd} Wei Su}
\IEEEauthorblockA{\textit{Lanzhou University}\\
Lanzhou, China \\
suwei@lzu.edu.cn}
\and
\IEEEauthorblockN{\textsuperscript{*} Jiamin Liu}
\IEEEauthorblockA{\textit{Lanzhou University}\\
Lanzhou, China \\
jmliu21@lzu.edu.cn}
\and
\IEEEauthorblockN{\textsuperscript{*} Wenli Yue}
\IEEEauthorblockA{\textit{Lanzhou University}\\
Lanzhou, China \\
yuewl21@lzu.edu.cn}
\thanks{* Equal contribution}
}
\maketitle
\begin{abstract}
    As the rapid development of Intelligent Tutoring Systems (ITS) in the past decade, tracing the students' knowledge state has become more and more important in order to provide individualized learning guidance. This is the main idea of \textit{Knowledge Tracing (KT)}, which models students' mastery of knowledge concepts (KCs, skills needed to solve a question) based on their past interactions on platforms. Plenty of KT models have been proposed and have shown remarkable performance recently. However, the majority of these models use concepts to index questions, which implies that the predefined skill tags for each question are required in advance to indicate the specific KCs needed for answering the question correctly. This makes it pretty hard to apply on large-scale online education platforms where questions are often not well-organized by skill tags. In this paper, we propose Q-matrix-based Attentive Knowledge Tracing (QAKT), an end-to-end KT model that utilizes the attentive approach in situations where predefined skill tags are not available. With a novel hybrid embedding method based on the q-matrix and Rasch model, QAKT is capable of modeling problems hierarchically and learning the q-matrix efficiently based on students' sequences. Meanwhile, the architecture of QAKT ensures that it is friendly to questions associated with multiple skills and has outstanding interpretability. After conducting experiments on a variety of open datasets, we empirically verified that even without predefined skill tags, our model performs similarly to or even better than the state-of-the-art KT methods, by up to 2\% in AUC in some cases. Moreover, our model outperforms existing models that do not require skill tags as well (by up to 7\% in AUC) in predicting future learner responses. Results of further experiments suggest that the q-matrix learned by QAKT is highly model-agnostic and more information-sufficient than the one labeled by human experts, which could help with the data mining tasks in existing ITSs.
\end{abstract}


\section{Introduction}
As the spread of the Internet, the shortcomings of traditional educational methods, such as inflexible teaching schedules and obsolete teaching materials, have become more and more intolerable. To improve the efficiency of learning, people resort to a new technique called Knowledge Tracing (KT), to track student proficiency based on their past interactions with online educational platforms \cite{Anderson1990}. After decades of development, plenty of KT models have been proposed and proven to be effective under certain circumstances \cite{Abdelrahman2022}.

Earlier in this region, people tended to model student proficiency in a simple and interpretable way. The most typical and popular one is, to the best of our knowledge, Bayesian Knowledge Tracing (BKT) \cite{Corbett1995}, which treats the learning procedure as a Markov chain parameterized by guessing, slipping, acquiring, and initial learning and deduces the probability of a student answering future questions correctly with latent variables in the Hidden Markov Model (HMM) \cite{Rabiner1989}. However, BKT assumes that once the student has learned the skill, he or she will never forget it in the following interactions, which is unrealistic \cite{Abdelrahman2022}.

Inspired by the success of deep learning \cite{AlexGraves2013,Vaswani2017}, recent developments in KT mostly focus on how to model student behavior with trainable parameters and how to optimize it based on deep learning. Plenty of these models have demonstrated their efficiency and have promising results in predicting future student responses \cite{Piech2015,Zhang2017,Minn2022,Ghosh2020}. Deep Knowledge Tracing (DKT) \cite{Piech2015} is the first model attempting to tackle the KT problem with a deep learning-based method. Leveraging recurrent neural networks \cite{Elman1990}, DKT models student knowledge states as proficiency on all of the predefined knowledge concepts (KCs, skills needed to solve a question) and performs much better than traditional KT methods. With the continuous development of deep learning technology, there are already many studies attempting to incorporate the attention mechanism into KT models, of which the most typical one is attentive knowledge tracing (AKT) \cite{Ghosh2020}. AKT encodes each interaction as a mixture of the KC embedding and the difficulty parameter of the question based on the Rasch model \cite{Rasch1981ProbabilisticMF}, after which it leverages attention mechanisms to extract the hidden knowledge state from past interaction encodings. According to the experimental results on common KT datasets in \cite{Ghosh2020}, AKT performs significantly better than previous methods (e.g., DKT) in predicting future student responses. Moreover, the computational efficiency and interpretability are much higher as well.

\begin{figure*}
\centerline{\includegraphics[width=\textwidth]{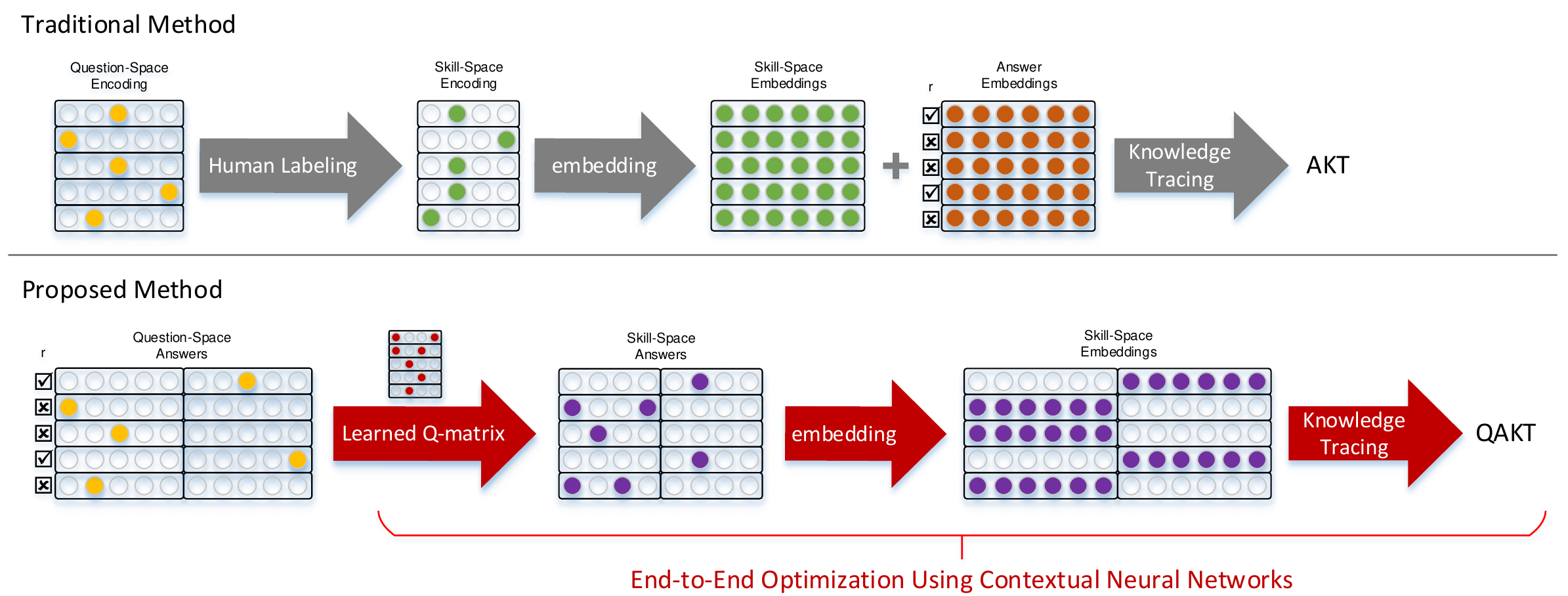}}
\caption{Overviews of the traditional method and the proposed method. AKT is trained with predefined skill tags by human experts, while QAKT is trained with skill tags labeled by itself instead of human experts.}
\label{fig_archis_comp}
\end{figure*}

However, all of the aforementioned models take KC sequences as input, requiring pre-defined skill tags by experts in advance and making the application on large-scale online educational platforms rather difficult. Besides, there are probably unpredictable biases and subjective tendencies in the skill tags labeled by experts \cite{Wang2022}. On the other hand, due to the architecture, these models can only accept one skill tag for each question, which is often not the case in reality \cite{Nakagawa2018}, resulting in a limitation on the application of these models \cite{Wang2022}.

In order to address this issue, an initial end-to-end KT solution is proposed: the end-to-end DKT (E2E-DKT) \cite{Nakagawa2018} model. This model attempts to learn the q-matrix automatically while fitting the students' past interaction sequences with a built-in Q-Embedding model. Hence, E2E-DKT is not only capable of running without human-defined skill tags, but it also achieves equal or even better results than DKT in predicting future responses on two public datasets \cite{Nakagawa2018}. Yet, E2E-DKT still suffers from some inherited disadvantages of DKT, such as low computational efficiency, poor predictive performance, and limited interpretability. Therefore, it cannot meet the demand for an efficient end-to-end KT model on current online education platforms.

In this paper, we propose a Q-matrix-based Attentive Knowledge Tracing (QAKT) method, the first attentive KT method that is able to learn the q-matrix from student interactions automatically. The fundamental insight underpinning our model is that questions and skills in attentive knowledge tracing are not necessarily mutually exclusive, and the skill labels associated with each question can be acquired through analysis of student responses, thereby enabling the automatic extraction of a q-matrix that contains sufficient information. Furthermore, the acquired q-matrix can be utilized as internal pre-trained features in any KT model, translating the interactions from question-space to low-dimensional skill-space. This facilitates the retraining of all other parameters, resulting in a well-performed KT model. The architecture of the traditional method (e.g., AKT) and proposed method is shown in Fig.~\ref{fig_archis_comp}.

QAKT relaxes the requirements for human-defined skill tags in KT, alleviates the problem of sparsity of exercises \cite{Piech2015}, and can therefore extend the application of KT in the real world. Using four benchmark real-world educational datasets, we empirically validated that our model can achieve similar or even better results than state-of-the-art models without requiring predefined tags by human experts. Furthermore, we conducted a comparative analysis of the efficacy of the q-matrix learned by our model, as opposed to the q-matrix obtained by E2E-DKT and the q-matrix labeled by human experts, in predicting future student responses. Our findings indicate that the q-matrix learned by our model is highly model-agnostic and better matches the characteristics of student sequences, thus making it capable of performing better in predicting future student responses\footnote{Source code and datasets will be available at \url{https://github.com/UnknownBen/QAKT}}.

The main contributions of this work are as follows:
\begin{itemize}
\item We propose Q-matrix-based Attentive Knowledge Tracing. Contrary to conventional KT methods, our model is able to automatically learn the q-matrix from student interactions with the help of attention mechanisms, without any predefined skill tags involved at all. Using four open datasets, we empirically validated that our model achieves similar or even better results in AUC compared to state-of-the-art KT models.
\item Motivated by discoveries pertaining to the q-matrix within the realm of cognitive diagnostic evaluation, we propose a straightforward, highly interpretable, yet very efficient interaction encoding method for KT based on the q-matrix and the Rasch model. Our method alleviates the issue of exercise data sparseness and improves the ability of the KT models to capture the relationships between different questions, thus making it possible to trace students' knowledge in an end-to-end manner.
\end{itemize}

\section{Related work}
\subsection{Attentive knowledge tracing}

Attentive knowledge tracing (AKT) \cite{Ghosh2020} is an attentive model that incorporates the monotonic attention mechanism and the Rasch model. Compared to common KT models (e.g., BKT, DKT), AKT performs significantly better in predicting future student responses and has higher computational efficiency \cite{Ghosh2020}. According to \cite{Ghosh2020}, the input of the AKT model mainly contains three parts of information: question, skill (labeled by human experts), and response. Due to the limitations of input format and model design, each interaction in AKT can only have one skill tag. This means that questions associated with multiple skills cannot be inputted into AKT directly. Typically, for each of these questions, we must pick one of its associated skill tags as its actual tag or simply combine all the associated skill tags into a brand new skill tag and assign the new skill tag to it. There will be inevitable information loss either way. 

AKT encodes the question at time step $t$ with input question and skill tag based on the Rasch model \cite{Rasch1981ProbabilisticMF} as follows: 
\begin{equation}
\mathbf{x}_t=\mathbf{c}_{c_t} + \mu_{q_t}\cdot \mathbf{d}_{c_t} \label{akt_xt_encode}
\end{equation}
where $c_t$ is the only skill associated with the current question $q_t$, $\mathbf{c}_{c_t}\in\mathbb{R}^D$ is the only embedding vector corresponding to skill $c_t$, $\mathbf{d}_{c_t}\in\mathbb{R}^D$ is a vector that summarizes the variation in questions covering $c_t$, and $\mu_{q_t}\in\mathbb{R}$ is a scalar difficulty parameter corresponding to question $q_t$, showing how far this question deviates from the concept it covers. The response of this interaction is encoded as follows:
\begin{equation}
\mathbf{y}_t=\mathbf{e}_{c_t, r_t}+\mu_{q_t}\cdot \mathbf{f}_{c_t, r_t}
\end{equation}
where $\mathbf{e}_{c_t, r_t}\in\mathbb{R}^D$ and $\mathbf{f}_{c_t, r_t}\in\mathbb{R}^D$ are the concept-response embedding and variation vectors, respectively. To be clear, $\mathbf{c}_{c_t}$, $\mathbf{d}_{c_t}$, $\mathbf{e}_{c_t, r_t}$, $\mathbf{f}_{c_t, r_t}$, and $\mu_{q_t}$ mentioned above are trainable parameters and independent of each other, reflecting features in different aspects of this interaction.

After encoding, AKT performs monotonic self-attention on $\mathbf{x}_t$ and $\mathbf{y}_t$, respectively, after which it performs the last monotonic attention computation to finally evaluate the student's knowledge state in the knowledge retriever. Based on intuition, \cite{Ghosh2020} proposed a context-aware distance measure for KT, together with a learnable decay rate parameter, to control the rate at which the attention weights decay as the distance between the current interaction and the previous interaction increases. Let $\mathbf{q}_t\in\mathbb{R}^{D_q}$, and $\mathbf{k}_t\in\mathbb{R}^{D_k}$ denote the query and key corresponding to the question the student responds to at time $t$, of which $D_q=D_k$, respectively. The context-aware distance between time $t$ and $\tau$, i.e., $d(t, \tau)$ is computed as follows:
\begin{align}
d(t, \tau) &=\vert{t-\tau}\vert \cdot \sum_{t^{\prime}=\tau+1}^t \gamma_{t, t^{\prime}} \label{akt_d}\\
\gamma_{t, t^{\prime}} &=\frac{exp(\frac{\mathbf{q}_t^T\mathbf{k}_{t^{\prime}}}{\sqrt{D_k}})} {\sum_{1\leq\tau^{\prime}\leq t}exp(\frac{\mathbf{q}_t^T\mathbf{k}_{\tau^{\prime}}}{\sqrt{D_k}})}
, \ \forall t^{\prime} \leq t \label{akt_gamma}.
\end{align}
where $\gamma_{t, t^{\prime}}$ is the importance indicator of the interaction at time $t^{\prime}$ ($\tau+1 \leq t^\prime \leq t$) to the one at time $t$. What is important, while performing self-attention on $\mathbf{x}_t, \mathbf{y}_t$ in the encoders, $\tau\leq t, \tau^{\prime}\leq t, t^{\prime}\leq t$. However, in the knowledge retriever, to prevent the model from cheating by peeking at the current response, $\tau < t$, $\tau^\prime<t$, and $t^\prime<t$ are ensured while processing.

The output of the knowledge retriever is concatenated with the current question embedding $\mathbf{x}_t$ after attention, then fed into a fully-connected network consisting of several fully-connected layers and an extra sigmoid function at the end to predict the correct probability $\hat{r}_t\in[0,1]$. The training objective of AKT is to minimize the binary cross-entropy loss of all learner responses:

\begin{equation}
L_p=\sum_i\sum_t \ell(\hat{r}_t^i, r_t^i) \label{akt_loss}
\end{equation}
where $r_t^i$ is the actual response for student $i$ at time $t$, 1 if the question is answered correctly, 0 otherwise. $\hat{r}_t^i$ is the predicted probability for student $i$ answering correctly at time $t$. $\ell$ is the binary cross entropy.

\subsection{Q-matrix} \label{q_matrix_sec}
The q-matrix \cite{Tatsuoka1983} is an important concept in the field of knowledge discovery \cite{DeLaTorre2008}, serving as a static matrix that delineates the correlation between a series of observable variables (i.e., questions in KT) and latent variables \cite{Barnes2005} (i.e., KCs in KT). For KT, the q-matrix describes the skills associated with each question in the dataset, making it possible to translate interactions from question-space to skill-space. Typically, the number of questions is far greater than the number of skills, thus the translation may yield advantages in mitigating the issue of exercise data sparseness. An example q-matrix is given in Table~\ref{q_matrix_exp}.

\begin{table}
\caption{Example q-matrix}
\label{q_matrix_exp}
\resizebox{\linewidth}{!}{
\begin{tabular}{*{6}{c}}
    \toprule
    & $q_1$ & $q_2$ & $q_3$ & $q_4$ & $q_5$ \\
    \midrule
    $c_1$ & \textbf{1} & \textbf{0} & \textbf{0} & \textbf{0} & \textbf{1}\\
    $c_2$ & \textbf{1} & \textbf{1} & \textbf{0} & \textbf{1} & \textbf{0}\\
    $c_3$ & \textbf{1} & \textbf{1} & \textbf{1} & \textbf{0} & \textbf{0}\\
    $c_4$ & \textbf{1} & \textbf{1} & \textbf{1} & \textbf{0} & \textbf{1}\\
    \bottomrule
\end{tabular}
}
\end{table}

In Table~\ref{q_matrix_exp}, each row corresponds to a unique skill, and each column corresponds to a unique question. Each value in the table, denoted by Q($c$, $q$), represents the probability of a student answering the question $q$ correctly, given that he or she has mastered all other skills in the table except skill $c$. Generally speaking, each value in a q-matrix is either 0 or 1, thus Q($c$, $q$) can also be interpreted as the necessity of mastering skill $c$ to correctly answer question $q$, 1 for true, 0 otherwise. For example, in order to answer question $q_2$ correctly, as stated in \ref{q_matrix_exp}, the student must master skills $\operatorname{c_2-c_4}$, while the mastery of skill $c_1$ is not required.

On small datasets (i.e., containing fewer than 100 questions and fewer than 30 skills), some scholars have attempted to extract the q-matrix through various algorithms based on phased feedback data \cite{Tiffany2005, Sun2014, Desmarais2012} or real-time feedback data \cite{Matsuda2015} as KT datasets \cite{Liu2019}, and achieved promising results. However, these algorithms are difficult to apply to datasets where there are a lot of questions or skills, which is often the case with online education platforms nowadays. Therefore, the common method to construct the q-matrix in KT is still defined by human experts, which is rather difficult to obtain while the database of questions keeps growing, and may contain biases or subjective tendencies. Hence, the extraction of the q-matrix in KT is still an open problem.

\subsection{End-to-End Deep knowledge tracing} \label{e2edkt}

Based on DKT, the end-to-end deep knowledge tracing (E2E-DKT) model \cite{Nakagawa2018} leverages recurrent neural networks (RNN) \cite{Williams1989} to learn the q-matrix from student sequences. Similar to DKT, two binary vectors indicating whether the question at time $t$ was answered correctly or incorrectly, respectively, are concatenated into the encoding in E2E-DKT, i.e., $\mathbf{x}_t$. Let $M$ denote the number of unique questions in the dataset, and $N$ denote the number of unique skills assumed in the dataset. The encoding for each interaction in E2E-DKT is a vector of length $2M$, in which the first half and the second half (hereinafter called $\textit{left\ vector}$ and $\textit{right\ vector}$) represent the encodings for the correct question and the incorrect question, respectively. According to the work in \cite{Piech2015}, this encoding method helps to improve the performance of KT models.

Furthermore, to translate the interaction between question-space and skill-space, E2E-DKT added two hidden layers: $\mathbf{u}_t$ and $\mathbf{v}_t$. The added feedforward layers before DKT can be reformulated as follows:
\begin{align}
\mathbf{P} &= \sigma(\mathbf{W}_{xu}) \\
\mathbf{x}_t &= [\mathbf{x}_t^{pos} \parallel \mathbf{x}_t^{neg}] \\
\mathbf{u}_t &= [\mathbf{P}\mathbf{x}_t^{pos} \parallel \mathbf{P}\mathbf{x}_t^{neg}]
\end{align}
where $\mathbf{x}_t^{pos}\in\mathbb{R}^M$ and $\mathbf{x}_t^{neg}\in\mathbb{R}^M$ represent the left and right vectors of the input encoding $\mathbf{x}_t$, respectively; $\parallel$ is the concatenation operation; $\mathbf{W}_{xu}$ is a weight matrix. $\mathbf{P}\in\mathbb{R}^{N\times M}$ is a table of the relevance between each question and each skill.

After translation from question-space to skill-space, E2E-DKT feeds $\mathbf{u}_t$ into a standard DKT, retrieving $\mathbf{v}_t$, which represents the current knowledge state in skill-space. Finally, it translates the knowledge state from skill-space to question-space, retrieving predicted probabilities of the student answering each question correctly at time $t+1$ as:
\begin{align}
\mathbf{y}_t &= \sigma(\mathbf{W}_{vy}\mathbf{v}_t + \mathbf{b}_y)
\end{align}
where $\mathbf{W}_{vy}$ is a weight matrix and $\mathbf{b}_y$ is a bias term.

The q-matrix mentioned in Section~\ref{q_matrix_sec} may be obtained from the relevance matrix as follows:
\begin{equation}
\mathbf{P}_{i, j}^{\prime} = 
\begin{cases}
1 & if \ \mathbf{P}_{i, j}=max(\mathbf{P}_i) \ or \ \mathbf{P}_{i, j} \geq \theta \\
0 & else
\end{cases}
\label{eq_e2edkt_binarization}
\end{equation}
where $i$ and $j$ are the indices of rows and columns of $\mathbf{P}$ (or $\mathbf{P}^{\prime}$), corresponding to a single skill or question, respectively. The threshold $\theta$ is a hyperparameter. 

During the training of the q-matrix, i.e., the training of the Q-Embedding model in E2E-DKT, \cite{Nakagawa2018} introduced two regularization techniques - reconstruction regularization loss and sparse regularization loss - to assess the quality of the learned q-matrix. The proposed regularization losses are defined as follows:
\begin{align}
\mathbf{x}^{\prime}_t &= \sigma(\mathbf{W}_{vy}\mathbf{u}_t^{pos} + \mathbf{b}_y) \\
L_r &= \sum_t \ell(\mathbf{x}^{\prime T}_t\delta(q_t), r_t) \\
L_s &= \sum_t (0.5 - \vert \mathbf{u}_t - 0.5 \vert)
\end{align}
where $\delta(q_{t+1})$ is the one-hot encoding of the question at time $t+1$.
The reconstruction regularization loss, $L_r$, indicates the intuition that the probability of a student answering the question correctly could be estimated based on his or her understanding of each concept in skill-space \cite{Nakagawa2018}. Hence, for a well-organized q-matrix, $L_r$ is relatively small. The sparse regularization loss, $L_s$, reflects the distance from values in $\mathbf{P}$ to 0 or 1. By minimizing $L_s$, the model is capable of suppressing information loss to a reasonable range while binarizing. The training objective is to minimize the weighted sum of $L_r$, $L_s$, and the original negative log likelihood in DKT \cite{Piech2015}.

\section{Proposed method}
\begin{figure*}
    \centerline{\includegraphics[width=\textwidth]{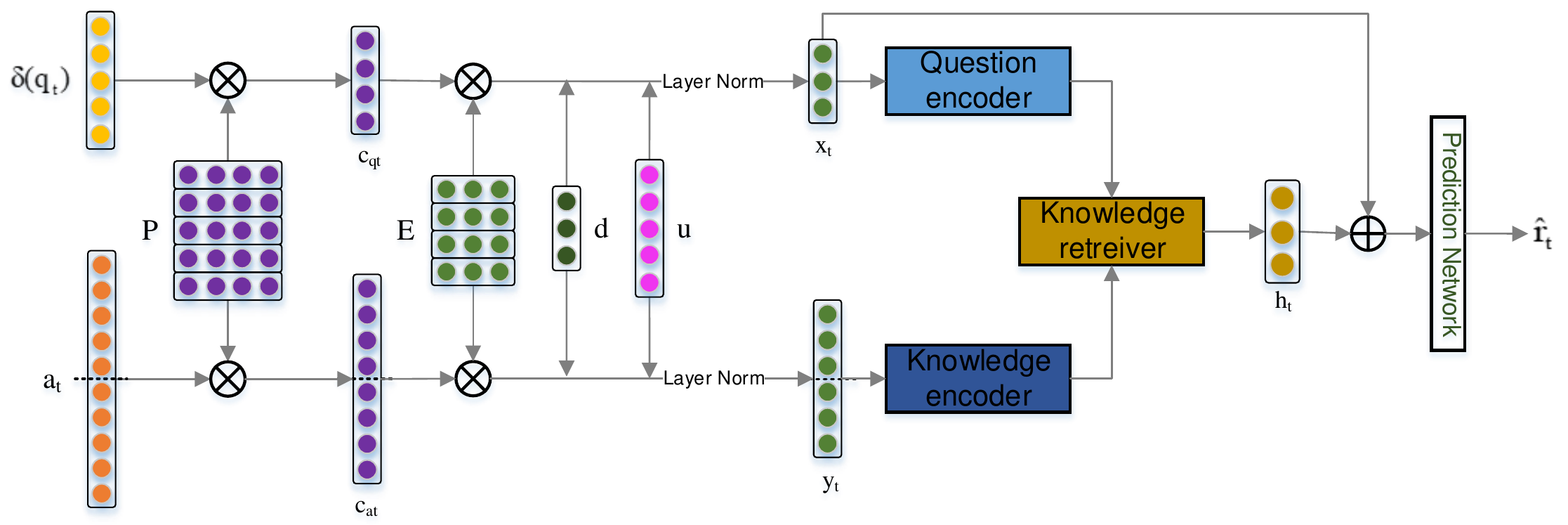}}
    \caption{Architecture overview of QAKT. QAKT embeds questions and responses based on the q-matrix and Rasch model first, after which it evaluates student proficiency on each skill at the current time step in the attention module and predicts the response in a fully-connected network. QAKT optimizes all the parameters inside, q-matrix included, by stochastic gradient descent in an end-to-end manner.}
    \label{qakt_full_archi}
\end{figure*}
In this section, we introduce the overall architecture of QAKT. Firstly, we illustrate the model architecture in cartoons and formulate the proposed interaction encoding method, which incorporates both the Rasch model and an initialized or pre-trained q-matrix. Then, we illustrate the attention module, response prediction network, and model optimization in mathematical equations, respectively. The architecture of QAKT is shown in Fig.~\ref{qakt_full_archi}.

\subsection{Exercise Representation}
Considering the characteristics of the students' learning process, we use a classic model in psychometrics, the Rasch model, to encode the interactions in KT. The Rasch model has demonstrated its efficacy in various related tasks by solely taking into account the complexity of the question and the learner's aptitude when assessing knowledge states. As denoted in Section~\ref{e2edkt}, $N$ is the number of unique skills assumed in the dataset, which may be set arbitrarily as a hyperparameter, and $M$ is the number of unique questions in the dataset. Taking the interaction at time $t$ as an example, the skill tags are extracted from the initialized or pre-trained q-matrix $\mathbf{P}$ by the following equations:
\begin{align}
\mathbf{P} &= \sigma(\mathbf{W}_p) \label{eq_qakt_P}\\
\mathbf{c}_{q_t} &= \mathbf{P}\delta(q_{t}) \label{eq_qakt_cqt}
\end{align}
where $\mathbf{P}\in\mathbb{R}^{N\times M}$ is a table of the relevance between individual questions and individual skills. Upon binarization, it can be converted into the q-matrix mentioned in Section~\ref{q_matrix_sec}. $\sigma(\cdot)$ is the sigmoid function. $\mathbf{W}_p\in\mathbb{R}^{N\times M}$ is a weight matrix. $\mathbf{c}_{q_t}\in\mathbb{R}^N$ is a vector containing the skill tags of question $q_t$, with $t>0$ and each value indicating the relevance of a particular skill to the current question: the larger the value, the more relevant the skill is.

Given the skill-tag vector $\mathbf{c}_{q_t}$, the skill encoding of this question is obtained by the following equation:
\begin{align}
\mathbf{k}_{q_t} &= \frac{ReLU(\mathbf{E}\mathbf{c}_{q_t} + \mathbf{d})}{\sum_j \mathbf{c}_{q_t}^j} \label{eq_qakt_kqt}
\end{align}
where $\mathbf{E}\in\mathbb{R}^{D\times N}$ is the skill embedding matrix, assigning an embedding vector of length $D$ for each assumed skill. $\mathbf{d}\in\mathbb{R}^D$ is the bias vector summarizing the deviation of the current skill encoding from the weighted sum of skill embeddings, indicating the variation in assumed skills. $\mathbf{c}_{q_t}^j$ is the value of index $j$ in $\mathbf{c}_{q_t}$.

Finally, we add the question difficulty parameter to the skill encoding and perform layer normalization \cite{Ba2016}, thus obtaining the embedding of the question $q_t$, i.e., $\mathbf{x}_t$:
\begin{align}
\mathbf{x}_t &= LayerNorm(\mathbf{k}_{q_t} + \mu_{q_t}) \label{eq_qakt_xt} 
\end{align}
where $\mu_{q_t}\in\mathbb{R}$ is a trainable scalar difficulty parameter in the vector $\mathbf{u}\in\mathbb{R}^{M}$, indicating how far this question deviates from the skill encoding to which it corresponds. We perform layer normalization at the end to improve training efficiency and ensure the effectiveness of the trained q-matrix.

For the response, similar to \cite{Piech2015}, we offset the one-hot vector of the question according to the actual response of the student first, thus obtaining the question-skill encoding, which is composed of two parts: $\mathbf{a}_t^{pos}$ and $\mathbf{a}_t^{neg}$. In our experiments, we observed that treating the zero part in question-skill encoding as a hypothesized padding question is much more effective, which means the skill tags of the padding question ($q_0$) are trainable. Furthermore, we extract the skill tags, mix and average the skill embeddings, introduce the current question difficulty for each part, respectively, and finally perform layer normalization. The equations are as follows:
\begin{small}
\begin{align}
\mathbf{a}_t &= [\mathbf{a}_t^{pos} \parallel \mathbf{a}_t^{neg}] \\
\mathbf{c}_{a_t} &= [\mathbf{P}\mathbf{a}_t^{pos} \parallel \mathbf{P}\mathbf{a}_t^{neg}] \label{eq_qakt_ct} \\
\mathbf{k}_{a_t} &= [\frac{ReLU(\mathbf{E}\mathbf{c}_{a_t}^{pos} + \mathbf{d})}{\sum_j {\mathbf{c}_{a_t}^{pos}}^j} \parallel \frac{ReLU(\mathbf{E}\mathbf{c}_{a_t}^{neg} + \mathbf{d})}{\sum_j {\mathbf{c}_{a_t}^{neg}}^j}] \label{eq_qakt_kat} \\
\mathbf{y}_t &= LayerNorm(\mathbf{k}_{a_t}  + \mu_{q_t}) \label{eq_qakt_yt}
\end{align}
\end{small}
where $\parallel$ represents concatenation. ${\mathbf{c}_{a_t}^{pos}}^j$ represents the value of index $j$ in the correct skill-tag vector $\mathbf{c}_{a_t}^{pos}$ at time step $t$. If the student answered the current question incorrectly, $\mathbf{c}_{a_t}^{pos}$ will be a vector of all zeros. ${\mathbf{c}_{a_t}^{neg}}^j$ is calculated in the same way.

After input encoding, each interaction is encoded into two corresponding parts: question embedding $\mathbf{x}_t\in\mathbb{R}^D$ and response embedding $\mathbf{y}_t\in\mathbb{R}^{2D}$. Let $l$ denote the length of the student sequence. The output of the embedding module is the question embedding matrix $\mathbf{X}\in\mathbb{R}^{D\times l}$ and response embedding matrix $\mathbf{Y}\in\mathbb{R}^{2D\times l}$, which help evaluate the knowledge states afterwards. The encoding results are as follows:

\begin{align}
\mathbf{X} &= \{\mathbf{x}_1, \mathbf{x}_2, ..., \mathbf{x}_l\}, \mathbf{x}_i\in\mathbb{R}^D & \\
\mathbf{Y} &= \{\mathbf{y}_1, \mathbf{y}_2, ..., \mathbf{y}_l\}, \mathbf{y}_i\in\mathbb{R}^{2D} &
\end{align}

\subsection{Attention Module}
Similar to AKT, we use monotonic attention mechanisms to evaluate the knowledge state according to the question embeddings and response embeddings obtained. The monotonic attention mechanism employed in QAKT can be reformulated as follows:

\begin{align}
Attention(Q, K, V) = Softmax(\frac{\mathbf{W}_{decay}\mathbf{Q}\mathbf{K}^T}{\sqrt {D_k}})\mathbf{V}
\label{att_eq}
\end{align}
where query-key pairs and values come from the embeddings:
\begin{align}
\mathbf{Q} = \mathbf{A}\mathbf{W}_Q, \mathbf{K} = \mathbf{A}\mathbf{W}_K, \mathbf{V} = \mathbf{B}\mathbf{W}_V
\end{align}
where $\mathbf{W}_Q$, $\mathbf{W}_K$, and $\mathbf{W}_V$ are the query, key, and value projection matrices used to project the embedding to different spaces, respectively, and they are all square matrices. $\mathbf{W}_{decay}\in\mathbb{R}^{l\times l}$ is a position encoding matrix with a decay term inside:
\begin{align}
\mathbf{W}_{decay} = e^{-\theta \mathbf{D}}
\end{align}
where $\theta > 0$ is a trainable decay rate parameter. Considering the performance of KT, we use the context-aware distance measure proposed in \cite{Ghosh2020}, and each value in the distance matrix $\mathbf{D}$ is defined by \eqref{akt_d}, \eqref{akt_gamma}.

There are three main parts in the attention module: a question encoder, a knowledge encoder, and a knowledge retriever. For the question encoder, we perform a self-attention mechanism on the question embedding matrix, which means setting $\mathbf{A}=\mathbf{B}=\mathbf{X}$, thus obtaining a more comprehensive question embedding matrix, i.e., $\mathbf{X^\prime}\in\mathbb{R}^{D\times l}$. The response embedding matrix with attentive information, i.e., $\mathbf{Y^\prime}\in\mathbb{R}^{2D\times l}$, is calculated in the same way. Then, we set $\mathbf{A}=\mathbf{X}^\prime$ and $\mathbf{B}=\mathbf{Y}^\prime$ for the knowledge retriever, performing the monotonic attention mechanism for the last time to evaluate the knowledge state of the student.

In the question encoder and knowledge encoder mentioned above, we allow the model to read the response of the current interaction for more comprehensive attentive information, i.e., $\tau\leq t, \tau^{\prime}\leq t, t^{\prime}\leq t$ in \eqref{akt_d} and \eqref{akt_gamma}. However, to prevent the model from cheating, access to the current response is not allowed in the knowledge retriever, i.e., $\tau < t, \tau^{\prime} < t, t^{\prime} < t$.

Among all three attention mechanisms of this module, to attend to information as comprehensively as possible, we perform attention computations in multiple heads ($n=8$ in this paper) and concatenate the results of all heads along the last dimension as the final attention output. To feed these heads, each one of the embedding matrices inputted is split into $n$ parts along the last dimension. The output of this module is $\mathbf{H}\in\mathbb{R}^{D\times l}$.

\subsection{Response Prediction Network}
The knowledge state extracted in the attention module is fed into the response prediction network, i.e., a fully-connected network consisting of three connected layers with decreasing dimensions in the last dimension (1 for the last layer), each of which is composed of a normalization layer, a fully-connected layer, and a dropout layer, to predict the final response of the interaction. The input of the network is a matrix concatenated by a knowledge state matrix $\mathbf{H}$ and a question embedding matrix $\mathbf{X}$. The concatenated question embedding is used to prompt the model with which question to answer. The output of the prediction network finally goes through the sigmoid function, thus obtaining the final prediction $\hat{r}_t\in[0, 1]$ of the interaction at time step $t, t\in\{1,2,...,l\}$.

\subsection{Optimization}
The key insight of our model is to train the q-matrix in an attentive way based on students' historical data, with the learned q-matrix meeting general requirements, e.g., each value in a q-matrix is either 0 or 1. Hence, the matrix of floating point numbers, $\mathbf{P}$, need to be binarized after the training in the first phase. In order to meet the sparsity requirement and suppress the information loss during binarization, we introduce the sparse reconstruction loss $L_s$ \cite{Nakagawa2018} in the loss function, defined as follows:

\begin{equation}
L_s = \sum_i\sum_t\sum_j{(0.5 - \vert \mathbf{c}_{q_t}^{i,j} - 0.5 \vert)}
\label{qakt_ls}
\end{equation}
where $\mathbf{c}_{q_t}^{i,j}$ is the value of index $j$ in the skill-tag vector of the question $q_t$ answered by student $i$. Moreover, in order to improve the effect of the question difficulty parameters $\mathbf{u}$, we add an L2 regularization term as follows:

\begin{equation}
L_c = \sum_{j=1}^M \mu_j^2
\end{equation}
where $\mu_j$ is the difficulty parameter of question $j$. The regularization terms mentioned above, in conjunction with the loss of AKT as described in \eqref{akt_loss}, may result in the total loss of QAKT as follows:

\begin{equation}
L = L_p + \beta L_s + \lambda L_c \label{loss_total}
\end{equation}
where $\beta$ and $\lambda$ are hyperparameters used to balance the proportion of the three parts of the loss.

\section{Experiments} \label{exp}
In this paper, we propose an attentive KT model that is able to learn the q-matrix from student interactions. Compared with conventional models that rely heavily on the skill tags labeled by human experts, QAKT is capable of modeling the features of questions with the attention mechanism and labeling the skill tags automatically, thus relaxing the requirements for KT model deployment. In this section, we conduct experiments on several benchmark educational datasets, evaluating the performance of QAKT and comparing it to state-of-the-art KT methods.

\subsection{Dataset}

For the experiment, we used four benchmark educational datasets: ASSISTments2009(ASSIST2009), ASSISTments2017(ASSIST2017), Statics2011\footnote{The ASSISTments2009, ASSISTments2017 and Statics2011 datasets are retrieved from \url{https://github.com/bigdata-ustc/EduData}}, and JunyiAcademy(Junyi)\footnote{The Junyi dataset is retrieved from \url{https://pslcdatashop.web.cmu.edu/DatasetInfo?datasetId=1198}}. The ASSISTments datasets were collected from an online tutoring platform---ASSISTments\footnote{https://www.assistments.org/}, in which the ASSIST2009 dataset has been the standard benchmark for KT methods over the last decade. The Statics2011 \cite{bigdata2021edudata} dataset was collected from the Engineering Statics course taught at Carnegie Mellon University during Fall 2011 \cite{Abdelrahman2022}. The Junyi dataset was collected from the online tutoring platform---Junyi Academy\footnote{https://www.junyiacademy.org/} in 2015\cite{Chang2015ModelingER}. For a fair comparison with models that require pre-defined skill tags (e.g., AKT), we eliminate interactions where there are null values in the student ID column, question column, skill column, or response column during preprocessing on all datasets. While training QAKT and E2E-DKT, we offer only student sequences of question IDs and responses. For models that lack the capability to handle questions associated to multiple skills (e.g., DKT, AKT), we input the lexicographically smallest skill tag among all skill tags associated with the given question as the actual skill tag of that question. For the sake of computational cost, we intercepted the first $1,000,000$ interactions on the Junyi dataset after arranging the data by the student ID and the timestamp of the interaction for the training of all models. We list the numbers of students, interactions, questions, and skill tags (labeled by human experts) of all datasets after preprocessing in Table~\ref{dataset_info}.

\begin{table}
\caption{dataset details}
\label{dataset_info}
\resizebox{\linewidth}{!}{
\begin{tabular}{*{5}{c}}
    \toprule
     & Statics2011 & ASSIST2009 & ASSIST2017 & Junyi \\
    \midrule
    Learners & 331 & 4,163 & 1,709 & 10,404 \\
    Skill tags & 85 & 123 & 102 & 39 \\
    Questions & 633 & 17,751 & 3,162 & 704 \\
    Responses & 111,298 & 338,001 & 938,371 & 999,995 \\
    \makecell[c]{Avg responses\\per question} & 175.83 & 19.04 & 296.77 & 1,420.45 \\
    \bottomrule
\end{tabular}
}
\end{table}

\subsection{Baseline methods and evaluation metrics}
We compare QAKT methods with several baseline KT methods in our experiments, DKT \cite{Piech2015}, dynamic key-value memory networks (DKVMN) \cite{Zhang2017}, AKT, E2E-DKT, and the recently proposed interpretable knowledge tracing (IKT) \cite{Minn2022} included. Among them, E2E-DKT takes exactly the same input as QAKT, i.e., sequences of questions and responses, and is able to learn the q-matrix automatically. The validity of the q-matrices learned in both methods will be analyzed later. Using external memory matrices for skill representations and knowledge state representations, DKVMN is a more interpretable method than DKT, which simply feeds raw skill tags into RNNs. IKT is a highly interpretable KT method because it defines the features (e.g., question difficulties, student ability profiles) in a statistical way by incorporating BKT and the Rasch model, thus having a relatively complex feature engineering procedure. IKT takes sequences of questions, skills, and responses as input and predicts the final response with a Tree Augmented Naive Bayes classifier (TAN) \cite{Friedman1997} after feature engineering. We use the area under the receiver operating characteristic curve (AUC) as the metric when evaluating the performance of all KT methods in predicting student responses.

\subsection{Implementation details}
The training objective of QAKT in the first phase is to obtain the q-matrix that best matches the sequence characteristics, thus labeling the questions automatically. Although the skill number $N$ may be defined as an arbitrary non-negative integer, for the sake of proper comparison with the q-matrix labeled by human experts, we set the skill number to the number of human-defined skills in the dataset, as shown in Table~\ref{dataset_info}. Moreover, to improve the efficiency of skill embeddings, we add a dropout layer with a drop rate of $0.05$ after each computation in which the skill embedding matrix $\mathbf{E}$ is involved. We implement all versions of QAKT in PyTorch \cite{pytorch}, so there are few trainable parameters in the layer normalization function.

During the training in the first phase, the hyperparameters in \eqref{loss_total} are set as $\beta=1, \lambda=10^{-5}$. After training in the first phase, the matrix $\mathbf{P}$ is binarized as follows to obtain the learned q-matrix $\mathbf{P}^{\prime}$:

\begin{equation}
\mathbf{P}_{i, j}^{\prime} = 
\begin{cases}
1 & if \ \mathbf{P}_{i, j} < \eta \cdot max(\mathbf{P}_i) \\
0 & else
\label{qakt_binarize}
\end{cases}
\end{equation}
where $\eta$ is a hyperparameter for binarization, set to $0.99$ in this paper. In the second phase, we fix the q-matrix learned in the first phase as $\mathbf{P}$ in \eqref{eq_qakt_P}, reinitialize other parameters, set $\beta=0, \lambda=10^{-5}$, and retrain other parameters in QAKT for the best model in predicting student responses.

To compare the model performance and effectiveness of the q-matrix learned by different models, we reimplement E2E-DKT\footnote{Source code will be available at \url{https://github.com/UnknownBen/E2E-DKT}} according to \cite{Nakagawa2018} in PyTorch. Specifically, GRU\cite{Bhaduri2014} is used as the sublayer $\varphi$ in E2E-DKT, and other hyperparameters are kept as consistent as possible with the original paper. We reimplement DKT in PyTorch, and use the implementation by the original authors for DKVMN, IKT, and AKT while keeping the hyperparameters unchanged. During the training of QAKT, AKT, and E2E-DKT, for the sake of computational cost, we cut student sequences longer than $200$ into multiple slices, set the batch size to $24$, the maximum number of iterations to $300$, and train them with the $Adam$ optimizer. For evaluation purposes, we perform standard k-fold cross-validation (with $k=5$) for all models on all datasets in this section. Hence, there are $20\%$ students in each one of the five folds. In each experiment, three folds are used as the training set, one fold is used as the validation set, and the fold left out is used for testing. We also perform early stopping based on the AUC metric on the validation set while training QAKT, AKT, and E2E-DKT. All the models are trained and evaluated on our machine, which is equipped with one GeForce GTX 1080 Ti GPU.

\section{Results and discussion}
In this section, we present the results of experiments, analyze and discuss the results, and conduct further experiments to verify the effectiveness of our model.

\subsection{Student Performance Prediction}
First, we compare the performance of QAKT and other models in predicting future student responses. Table~\ref{auc_tab} shows the average AUC metrics over five test sets for all KT methods on all datasets, in which the best performance on each dataset is shown in bold.

\begin{table}
\caption{Prediction performance of KT methods}
\label{auc_tab}
\resizebox{\linewidth}{!}{
\begin{tabular}{*{7}{c}}
\toprule
\multirow{2}{*}{\textbf{Dataset}} &\multicolumn{6}{c}{\textbf{AUC}} \\
\cline{2-7}
 & \textbf{\textit{DKT}}& \textbf{\textit{E2E-DKT}}& \textbf{\textit{DKVMN}}& \textbf{\textit{IKT}}& \textbf{\textit{AKT-R}}& \textbf{\textit{QAKT}} \\
\midrule
Statics2011& 0.7659 & 0.8062 & 0.8064 & 0.7802 & 0.8115 & \textbf{0.8209} \\
ASSIST2009& 0.8033 & 0.7475 & 0.8045 & 0.7408 & \textbf{0.8216} & 0.8171 \\
ASSIST2017& 0.6940 & 0.7690 & 0.6999 & 0.6992 & 0.7521 & \textbf{0.7739} \\
Junyi& 0.7443 & 0.7702 & 0.7476 & 0.7496 & 0.7867 & \textbf{0.7908} \\
\bottomrule
\multicolumn{7}{l}{$^{\mathrm{1}}$\textbf{Bold} numbers are the best performance.}
\end{tabular}
}
\end{table}

From the results in Table~\ref{auc_tab}, we observe that even though both are trained without predefined skill tags, QAKT (sometimes significantly) outperforms E2E-DKT by $1.5\%, 7.0\%, 0.5\%$, and $2.1\%$ in terms of AUC on the Statics2011, ASSIST2009, ASSIST2017, and Junyi datasets, respectively. The results suggest that the q-matrix learned by QAKT better matches the characteristics of student sequences than the one learned by E2E-DKT, leading to smaller labeling errors and better performance in predicting future student responses. Compared to models requiring predefined skill tags, we see that on the Statics2011, ASSIST2017, and Junyi datasets, QAKT outperforms all other KT models and achieves an AUC improvement over the nearest model (i.e., AKT-R) by a margin of $0.9\%$, $2.2\%$, and $0.4\%$, respectively. On the ASSIST2009 dataset, which has the fewest average responses per question, AKT-R slightly outperforms QAKT by $0.5\%$ in terms of AUC. Given that DKT marginally outperforms E2E-DKT by $5.6\%$ on the ASSIST2009 dataset in our experiments, we hypothesize that the reason is that the lack of interactions on questions results in larger labeling errors, thus affecting the effectiveness of the learned q-matrix for both methods. In general, without predefined skill tags, QAKT performs similarly or even better than state-of-the-art models by learning the q-matrix itself in KT.

We also notice that AKT-R performs significantly better than IKT on all datasets. Moreover, despite training without predefined skill tags, E2E-DKT performs significantly better than IKT on all datasets as well.

\subsection{Ablation Study} \label{sec_ablation_study}
To validate the key innovations of our encoding method for KT interactions, which include the activation function used for skill encoding, scaling operations based on skill-tag vectors, and layer normalization before the attention mechanism, we conducted additional ablation experiments to compare the performance of the original QAKT method with several variants that differ in the logic of equations \eqref{eq_qakt_xt} and \eqref{eq_qakt_yt} on three datasets: Statics2011, ASSIST2009, and ASSIST2017. We chose not to conduct ablation experiments on the Junyi dataset due to its limited number of skills and abundance of interactions, which may lead to biased results in the ablation studies. For the sake of proper comparison, all the models in this section are trained without early stopping, and other hyperparameters are kept the same as in Section \ref{exp}. The variants of the QAKT method compared in this section are as follows:
\begin{itemize}
\item QAKT-NoAct: This variant includes all the features of QAKT except for the activation function on the weighted sum of skill embeddings.
\item QAKT-NoAvg: In this variant of the default architecture, the scaling operation based on the element sum of its skill-tag vector $\mathbf{c_{q_t}}$ is removed.
\item QAKT-NoLN: In this variant of the default architecture, the layer normalization operation on exercise encoding before performing self-attention is removed.
\end{itemize}
Table \ref{ablation_tab} summarizes the average AUC results for different variants of the QAKT model, in which the best performances are shown in bold. We see that QAKT outperforms its counterparts, QAKT-NoAct, QAKT-NoAvg, and QAKT-NoLN, on all datasets. The experimental results suggests that: the non-linear transformation introduced by the activation function facilitates the modeling of skills, thus helping to solve more complex tasks, i.e., accurately assessing knowledge states in this case; the scaling operation while mixing skill embeddings helps to balance the different embeddings of skills associated with the question; the layer normalization before self-attention enhances the standardization of the final encodings. As a result, highly interpretable as it is, the proposed hierarchical encoding method helps alleviate the issue of exercise data sparseness and improves the performance of interaction encoding in KT. Most importantly, all the operations within it are necessary for an efficient KT model.

\begin{table}
\caption{Ablation Study}
\label{ablation_tab}
\resizebox{\linewidth}{!}{
\begin{tabular}{*{5}{c}}
\toprule
\multirow{2}{*}{\textbf{Dataset}} &\multicolumn{4}{c}{\textbf{AUC}} \\
\cline{2-5} 
 & \textbf{\textit{QAKT-NoAct}}& \textbf{\textit{QAKT-NoAvg}}& \textbf{\textit{QAKT-NoLN}}& \textbf{\textit{QAKT}} \\
\midrule
Statics2011& 0.8191 & 0.8208 & 0.8203 & \textbf{0.8212} \\
ASSIST2009& 0.8126 & 0.8157 & 0.7916 & \textbf{0.8163} \\
ASSIST2017& 0.7730 & 0.7727 & 0.7617 & \textbf{0.7739} \\
\bottomrule
\multicolumn{5}{l}{$^{\mathrm{1}}$\textbf{Bold} numbers are the best performance.}
\end{tabular}
}
\end{table}

\subsection{Skill Number Exploration} \label{sec_skill_number}

To investigate the impact of skill space size on KT performance, we evaluated our model in further experiments using varying numbers of skills. To ensure the validity of the evaluation and prevent dropout from affecting the size of effective skills, we removed the dropout layer from the exercise representation. Meanwhile, all the models in this section are trained without early stopping, while other hyperparameters are kept the same as in Section \ref{exp}. We conducted experiments on two datasets, Statics2011 and ASSIST2017, for two reasons. Firstly, they have a reasonable number of interactions averaged across questions, making it highly possible to construct the q-matrix precisely from these interactions. Secondly, there are few interactions in these datasets, which makes them a cost-effective reference for constructing the q-matrix.

\begin{figure}
    \includegraphics[width=\linewidth]{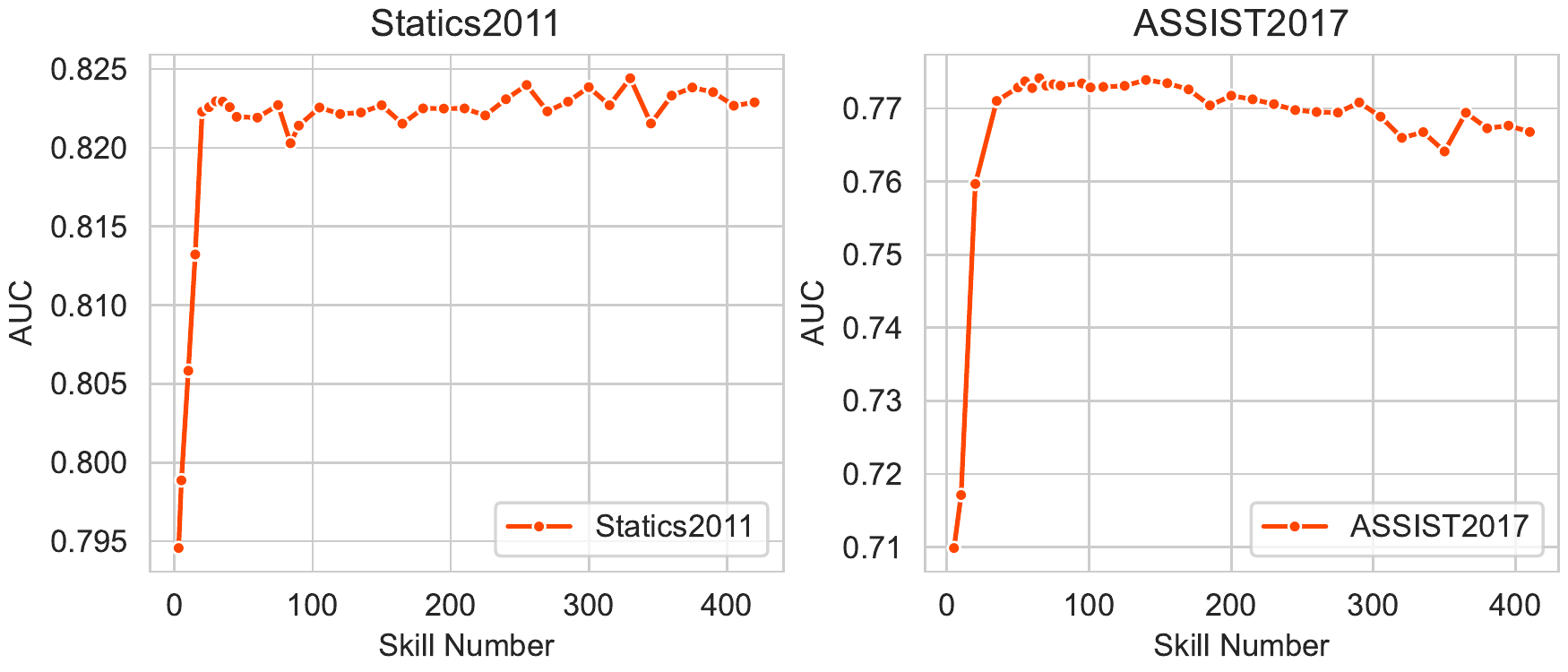}
    \caption{The AUC curve of QAKT with different skill numbers on the Statics2011 and ASSIST2017 datasets, without the use of early stopping or the dropout layer in the exercise representation. We observe a minimum threshold of optimal skill numbers on both datasets.}
    \label{qakt_test_skill_number}
\end{figure}

Fig.~\ref{qakt_test_skill_number} shows the results of experiments conducted on the Statics2011 and ASSIST2017 datasets with varying numbers of skills. We observe that on ASSIST2017, the models perform relatively better and are more stable when their skill number falls between 50 and 155, with the number labeled by experts being 102. The performance of models with a skill number below 50 or above 155 is relatively low, which suggests that the encoding in such a skill space is inadequate or redundant to distinguish between different questions in this dataset. On the Statics2011 dataset, we see that there is also a minimum threshold number (i.e., 20) of skills required to efficiently encode the questions. As a result, it appears that there is a minimum number of skills required for an efficient skill space for each dataset, which may be influenced by specific attributes of the dataset, such as the characteristics of the discipline and the size of the dataset. As the number of skills exceeds the minimum threshold within a certain range, the performance of our model in predicting future responses exhibits a gradual and consistent fluctuation, which suggests that our model is capable of adjusting the number of skills in some way. The threshold discovered in each dataset may help compress data and reduce the deployment costs of intelligent tutoring systems (ITS), thereby promoting the use of KT algorithms in various applications.

\subsection{Computational Efficiency} \label{sec_compu_cost}

\begin{figure*}
    \includegraphics[width=\textwidth]{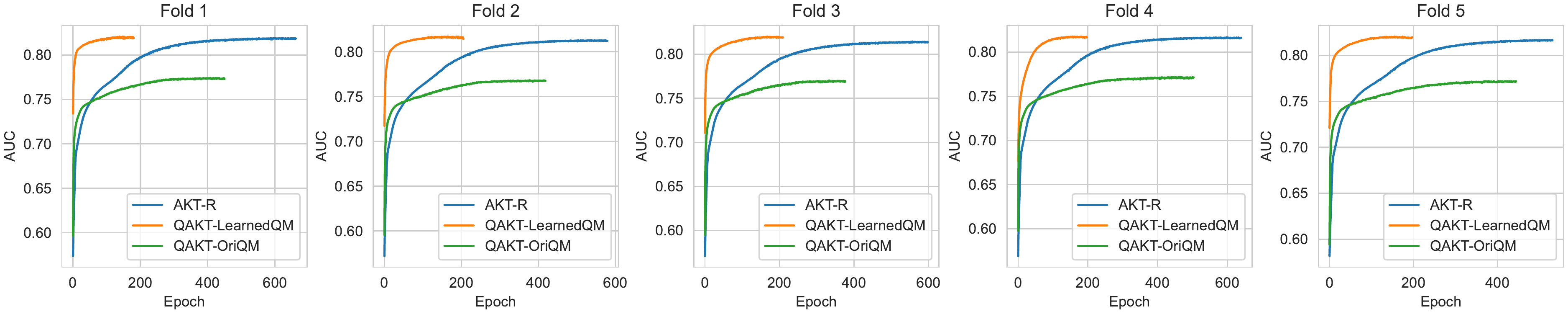}
    \caption{The AUC curve of QAKT with a self-trained q-matrix, QAKT with a human-labeled q-matrix, and AKT-R on the Statics2011 dataset, with the maximum number of iterations set to 1000. With the self-trained q-matrix, QAKT performs much better than that with the human-labeled q-matrix in terms of AUC on the Statics2011 dataset. On the other hand, without predefined skill tags, QAKT still performs slightly better than AKT-R on response prediction, and it converges faster as well.}
    \label{qakt_akt_statics2011_comp}
\end{figure*}

To evaluate the computational efficiency, we feed our model with q-matrices from different sources, fix them as the parameter matrix $\mathbf{P}$, and compare the performance in predicting future student responses with the original AKT-R model on the Statics2011 dataset. This dataset was selected for the same reasons as outlined in Section \ref{sec_skill_number}. The variants of the KT model in this experiment are as follows:

\begin{itemize}
\item QAKT-OriQM: The QAKT model with a fixed q-matrix labeled by human experts, i.e., extracted from the dataset. Because we already have the q-matrix, the model goes directly to the second phase of training. All skill tags associated with each question are preserved.
\item QAKT-LearnedQM: The QAKT model with a self-trained q-matrix in the first phase, goes directly to the second phase of training.
\item AKT-R: The original AKT-R model, which takes sequences of question, skill tag, and response as input. Similar to QAKT-OriQM, the skill tags are extracted from the dataset and created by human experts. Due to the limitation of the input format of AKT-R, questions with multiple skill tags are labeled with the one whose skill description (or skill ID) has the smallest lexicographical order.
\end{itemize}
All of the model variants above are trained on the same training set, and evaluated on the same validation set. The AUC metrics for each of the five experiments are presented in Fig.~\ref{qakt_akt_statics2011_comp}. For the integrity of the curve, we set the maximum number of iterations to 1000 for all three models.

The results in Fig.~\ref{qakt_akt_statics2011_comp} suggest that with the learned q-matrix in the first phase, QAKT not only achieves similar or even better performance than AKT-R, which heavily relies on human-labeled skill tags, but also converges faster. On the other hand, with the q-matrix learned by itself, QAKT performs much better than that with the human-labeled q-matrix on response prediction. It can be reasonably inferred that the labeling error of the q-matrix learned by QAKT is smaller than the one labeled by human experts, and thus the former one better matches the characteristics of student sequences. Moreover, the architecture of QAKT determines that it is more friendly to questions associated with multiple skills.

\subsection{Q-matrix Efficiency}
Due to the nature of q-matrix \cite{Tatsuoka2000}, it is only related to the dataset itself and not to the method of acquisition. In order to evaluate the effectiveness of the q-matrix learned by QAKT, we conducted further experiments on the Statics2011 and ASSIST2017 datasets. The reasons we selected these two datasets are that, in addition to the two given in Section~\ref{sec_skill_number}, E2E-DKT is a RNN-based model, thus training on datasets with too many questions (i.e., ASSIST2009) or too many interactions (i.e., Junyi) costs too much time according to our observations. For a fair comparison, q-matrices from three different sources were used in our experiments:

\begin{itemize}
\item $Q_{e2edkt}$: The q-matrix learned by E2E-DKT in the Q-Embedding Model.
\item $Q_{qakt}$: The q-matrix learned by QAKT in the first phase.
\item $Q_{ori}$: The q-matrix labeled by human experts, is extracted from the dataset. All skill tags associated with each question are preserved.
\end{itemize}
We feed the three q-matrices into E2E-DKT and QAKT, respectively, retrain other parameters, and evaluate the final performance on response prediction for each model on test sets. The q-matrix inputted is fixed as the parameter matrix $\mathbf{P}$ in both models. Moreover, we set $\beta=0, \lambda=10^{-5}$ for QAKT. In E2E-DKT, we train the model based on the given q-matrix directly, with the negative log-likelihood as the total loss function according to \cite{Nakagawa2018}. Other hyperparameters and settings are the same as in Section~\ref{exp}.

\begin{table}
\caption{Q-matrix Performance comparison}
\label{qm_auc_tab}
\begin{tabular}{*{5}{c}}
\toprule
\multirow{2}{*}{Tags} &\multicolumn{2}{c}{\textbf{Statics2011}} & \multicolumn{2}{c}{\textbf{ASSIST2017}}\\
\cline{2-3}
\cline{4-5}
& E2E-DKT & QAKT & E2E-DKT & QAKT \\
\midrule
\textbf{\textit{Existing tags}}& 0.8045 & 0.7722 & 0.7647 & 0.7299 \\ 
\textbf{\textit{E2E-DKT learned}} & 0.8062 & 0.8148 & 0.7690 & 0.7691 \\
\textbf{\textit{QAKT learned}} & \textbf{0.8133} & \textbf{0.8209} & \textbf{0.7725} & \textbf{0.7739} \\
\bottomrule
\multicolumn{5}{l}{$^{\mathrm{1}}$\textbf{Bold} numbers are the best performance.}
\end{tabular}
\end{table}

The results are presented in Table~\ref{qm_auc_tab}, in which all the performances are averaged over five experiments with different test sets. We see that on the Statics2011 dataset, E2E-DKT with $Q_{qakt}$ outperforms the same model with $Q_{e2edkt}$ and $Q_{ori}$ by a margin of $0.7\%$, and $0.9\%$ in AUC. On the other hand, QAKT with $Q_{qakt}$ performs better than that with $Q_{e2edkt}$ and $Q_{ori}$ by $0.6\%$, and $4.9\%$. On the ASSIST2017 dataset, compared with the same model with $Q_{e2edkt}$ and $Q_{ori}$, E2E-DKT with $Q_{qakt}$ improves the AUC by $0.3\%$, and $0.8\%$ while QAKT with $Q_{qakt}$ improves the AUC by $0.5\%$, and $4.4\%$. Taken together, the q-matrix learned by QAKT is highly independent and model-agnostic, fits student response sequences well no matter what model it is used in, and thus has great potential for application.

\section{Conclusion and future work}
In this paper, we propose a q-matrix-based attentive KT model (QAKT), which is able to learn the q-matrix from student interactions with attention mechanisms automatically. Our method relieves the issue of exercise data sparseness by building hierarchical representations based on the q-matrix and Rasch model, and it is adept at modeling the relationships between questions and skills. None of the existing KT methods is, to the best of our knowledge, able to learn the q-matrix with attention mechanisms. Using four open datasets, we empirically validated that, without predefined skill tags, our method achieves similar or even better performance than state-of-the-art KT methods, and it converges faster. Further experimental results show that the q-matrix learned by our model is more in line with the characteristics of student interactions than the q-matrix from other sources and exhibits excellent independence from the model it is learned by, thus having great potential for application on large-scale online education platforms.

As part of future work, we plan to investigate the relationship between skills in the learned q-matrix with graph attention networks, attempting to improve the interpretability of end-to-end KT models.

\section{Acknowledgment}
The work described in this paper was partially supported by the Supercomputing Center of Lanzhou University.
\normalem
\footnotesize
\bibliographystyle{IEEEtran}
\bibliography{refs}

\begin{thebibliography}{10}
\providecommand{\url}[1]{#1}
\csname url@samestyle\endcsname
\providecommand{\newblock}{\relax}
\providecommand{\bibinfo}[2]{#2}
\providecommand{\BIBentrySTDinterwordspacing}{\spaceskip=0pt\relax}
\providecommand{\BIBentryALTinterwordstretchfactor}{4}
\providecommand{\BIBentryALTinterwordspacing}{\spaceskip=\fontdimen2\font plus
\BIBentryALTinterwordstretchfactor\fontdimen3\font minus
  \fontdimen4\font\relax}
\providecommand{\BIBforeignlanguage}[2]{{%
\expandafter\ifx\csname l@#1\endcsname\relax
\typeout{** WARNING: IEEEtran.bst: No hyphenation pattern has been}%
\typeout{** loaded for the language `#1'. Using the pattern for}%
\typeout{** the default language instead.}%
\else
\language=\csname l@#1\endcsname
\fi
#2}}
\providecommand{\BIBdecl}{\relax}
\BIBdecl

\bibitem{Anderson1990}
\BIBentryALTinterwordspacing
J.~R. Anderson, C.~Boyle, A.~T. Corbett, and M.~W. Lewis, ``Cognitive modeling
  and intelligent tutoring,'' \emph{Artificial Intelligence}, vol.~42, no.~1,
  pp. 7--49, 1990. [Online]. Available:
  \url{https://www.sciencedirect.com/science/article/pii/000437029090093F}
\BIBentrySTDinterwordspacing

\bibitem{Abdelrahman2022}
\BIBentryALTinterwordspacing
G.~Abdelrahman, Q.~Wang, and B.~Nunes, ``Knowledge tracing: A survey,''
  \emph{ACM Comput. Surv.}, vol.~55, no.~11, feb 2023. [Online]. Available:
  \url{https://doi.org/10.1145/3569576}
\BIBentrySTDinterwordspacing

\bibitem{Corbett1995}
A.~T. Corbett and J.~R. Anderson, ``Knowledge tracing: Modeling the acquisition
  of procedural knowledge,'' \emph{User modeling and user-adapted interaction},
  vol.~4, pp. 253--278, 1994.

\bibitem{Rabiner1989}
L.~Rabiner, ``A tutorial on hidden markov models and selected applications in
  speech recognition,'' \emph{Proceedings of the IEEE}, vol.~77, no.~2, pp.
  257--286, 1989.

\bibitem{AlexGraves2013}
A.~Graves, A.-r. Mohamed, and G.~Hinton, ``Speech recognition with deep
  recurrent neural networks,'' in \emph{2013 IEEE International Conference on
  Acoustics, Speech and Signal Processing}, 2013, pp. 6645--6649.

\bibitem{Vaswani2017}
A.~Vaswani, N.~Shazeer, N.~Parmar, J.~Uszkoreit, L.~Jones, A.~N. Gomez,
  {\L}.~Kaiser, and I.~Polosukhin, ``Attention is all you need,''
  \emph{Advances in neural information processing systems}, vol.~30, 2017.

\bibitem{Piech2015}
C.~Piech, J.~Bassen, J.~Huang, S.~Ganguli, M.~Sahami, L.~Guibas, and
  J.~Sohl-Dickstein, ``Deep knowledge tracing,'' in \emph{Proceedings of the
  28th International Conference on Neural Information Processing Systems -
  Volume 1}, ser. NIPS'15.\hskip 1em plus 0.5em minus 0.4em\relax Cambridge,
  MA, USA: MIT Press, 2015, p. 505–513.

\bibitem{Zhang2017}
J.~Zhang, X.~Shi, I.~King, and D.-Y. Yeung, ``Dynamic key-value memory networks
  for knowledge tracing,'' in \emph{Proceedings of the 26th international
  conference on World Wide Web}, 2017, pp. 765--774.

\bibitem{Minn2022}
S.~Minn, J.-J. Vie, K.~Takeuchi, H.~Kashima, and F.~Zhu, ``Interpretable
  knowledge tracing: Simple and efficient student modeling with causal
  relations,'' in \emph{Proceedings of the AAAI Conference on Artificial
  Intelligence}, vol.~36, no.~11, 2022, pp. 12\,810--12\,818.

\bibitem{Ghosh2020}
A.~Ghosh, N.~Heffernan, and A.~S. Lan, ``Context-aware attentive knowledge
  tracing,'' in \emph{Proceedings of the 26th ACM SIGKDD international
  conference on knowledge discovery \& data mining}, 2020, pp. 2330--2339.

\bibitem{Elman1990}
\BIBentryALTinterwordspacing
J.~L. Elman, ``Finding structure in time,'' \emph{Cognitive Science}, vol.~14,
  no.~2, pp. 179--211, 1990. [Online]. Available:
  \url{https://www.sciencedirect.com/science/article/pii/036402139090002E}
\BIBentrySTDinterwordspacing

\bibitem{Rasch1981ProbabilisticMF}
G.~Rasch, ``Probabilistic models for some intelligence and attainment tests,''
  \emph{The SAGE Encyclopedia of Research Design}, 1981.

\bibitem{Wang2022}
\BIBentryALTinterwordspacing
W.~Wang, H.~Ma, Y.~Zhao, Z.~Li, and X.~He, ``Tracking knowledge proficiency of
  students with calibrated q-matrix,'' \emph{Expert Systems with Applications},
  vol. 192, p. 116454, 2022. [Online]. Available:
  \url{https://www.sciencedirect.com/science/article/pii/S0957417421017383}
\BIBentrySTDinterwordspacing

\bibitem{Nakagawa2018}
H.~Nakagawa, Y.~Iwasawa, and Y.~Matsuo, ``End-to-end deep knowledge tracing by
  learning binary question-embedding,'' in \emph{2018 IEEE International
  Conference on Data Mining Workshops (ICDMW)}.\hskip 1em plus 0.5em minus
  0.4em\relax IEEE, 2018, pp. 334--342.

\bibitem{Tatsuoka1983}
K.~K. Tatsuoka, ``Rule space: An approach for dealing with misconceptions based
  on item response theory,'' \emph{Journal of Educational Measurement},
  vol.~20, pp. 345--354, 1983.

\bibitem{DeLaTorre2008}
\BIBentryALTinterwordspacing
J.~De~La~Torre, ``An empirically based method of q-matrix validation for the
  dina model: Development and applications,'' \emph{Journal of Educational
  Measurement}, vol.~45, no.~4, pp. 343--362, 2008. [Online]. Available:
  \url{https://onlinelibrary.wiley.com/doi/abs/10.1111/j.1745-3984.2008.00069.x}
\BIBentrySTDinterwordspacing

\bibitem{Barnes2005}
T.~Barnes, ``The q-matrix method: Mining student response data for knowledge,''
  in \emph{American association for artificial intelligence 2005 educational
  data mining workshop}.\hskip 1em plus 0.5em minus 0.4em\relax AAAI Press,
  Pittsburgh, PA, USA, 2005, pp. 1--8.

\bibitem{Tiffany2005}
T.~Barnes, D.~Bitzer, and M.~Vouk, ``Experimental analysis of the q-matrix
  method in knowledge discovery,'' in \emph{Foundations of Intelligent
  Systems}, M.-S. Hacid, N.~V. Murray, Z.~W. Ra{\'{s}}, and S.~Tsumoto,
  Eds.\hskip 1em plus 0.5em minus 0.4em\relax Berlin, Heidelberg: Springer
  Berlin Heidelberg, 2005, pp. 603--611.

\bibitem{Sun2014}
Y.~Sun, S.~Ye, S.~Inoue, and Y.~Sun, ``Alternating recursive method for
  q-matrix learning,'' in \emph{Educational Data Mining}, 2014.

\bibitem{Desmarais2012}
\BIBentryALTinterwordspacing
M.~C. Desmarais, ``Mapping question items to skills with non-negative matrix
  factorization,'' \emph{SIGKDD Explor. Newsl.}, vol.~13, no.~2, p. 30–36,
  may 2012. [Online]. Available: \url{https://doi.org/10.1145/2207243.2207248}
\BIBentrySTDinterwordspacing

\bibitem{Matsuda2015}
N.~Matsuda, T.~Furukawa, N.~L. Bier, and C.~Faloutsos, ``Machine beats experts:
  Automatic discovery of skill models for data-driven online courseware
  refinement,'' in \emph{Educational Data Mining}, 2015.

\bibitem{Liu2019}
\BIBentryALTinterwordspacing
W.~P.-w. Y.~G. LIU Heng-yu, ZHANG Tian-cheng, ``A review of knowledge
  tracking,'' \emph{Journal of East China Normal University(Natural Science)},
  vol. 2019, no.~5, p.~1, 2019. [Online]. Available:
  \url{https://xblk.ecnu.edu.cn/EN/abstract/article_25656.shtml}
\BIBentrySTDinterwordspacing

\bibitem{Williams1989}
R.~J. Williams and D.~Zipser, ``A learning algorithm for continually running
  fully recurrent neural networks,'' \emph{Neural Computation}, vol.~1, no.~2,
  pp. 270--280, 1989.

\bibitem{Ba2016}
J.~Ba, J.~R. Kiros, and G.~E. Hinton, ``Layer normalization,'' \emph{ArXiv},
  vol. abs/1607.06450, 2016.

\bibitem{bigdata2021edudata}
bigdata ustc, ``Edudata,'' \url{https://github.com/bigdata-ustc/EduData}, 2021.

\bibitem{Chang2015ModelingER}
H.-S. Chang, H.-J. Hsu, and K.-T. Chen, ``Modeling exercise relationships in
  e-learning: A unified approach,'' in \emph{Educational Data Mining}, 2015.

\bibitem{Friedman1997}
\BIBentryALTinterwordspacing
N.~Friedman, D.~Geiger, and M.~Goldszmidt, ``Bayesian network classifiers,''
  \emph{Mach. Learn.}, vol.~29, no. 2–3, p. 131–163, nov 1997. [Online].
  Available: \url{https://doi.org/10.1023/A:1007465528199}
\BIBentrySTDinterwordspacing

\bibitem{pytorch}
A.~Paszke, S.~Gross, F.~Massa, A.~Lerer, J.~Bradbury, G.~Chanan, T.~Killeen,
  Z.~Lin, N.~Gimelshein, L.~Antiga, A.~Desmaison, A.~K\"{o}pf, E.~Yang,
  Z.~DeVito, M.~Raison, A.~Tejani, S.~Chilamkurthy, B.~Steiner, L.~Fang,
  J.~Bai, and S.~Chintala, \emph{PyTorch: An Imperative Style, High-Performance
  Deep Learning Library}.\hskip 1em plus 0.5em minus 0.4em\relax Red Hook, NY,
  USA: Curran Associates Inc., 2019.

\bibitem{Bhaduri2014}
\BIBentryALTinterwordspacing
K.~Cho, B.~van Merri{\"e}nboer, C.~Gulcehre, D.~Bahdanau, F.~Bougares,
  H.~Schwenk, and Y.~Bengio, ``Learning phrase representations using {RNN}
  encoder{--}decoder for statistical machine translation,'' in
  \emph{Proceedings of the 2014 Conference on Empirical Methods in Natural
  Language Processing ({EMNLP})}.\hskip 1em plus 0.5em minus 0.4em\relax Doha,
  Qatar: Association for Computational Linguistics, Oct. 2014, pp. 1724--1734.
  [Online]. Available: \url{https://aclanthology.org/D14-1179}
\BIBentrySTDinterwordspacing

\bibitem{Tatsuoka2000}
K.~K. Tatsuoka and G.~M. Boodoo, ``Subgroup differences on the gre quantitative
  test based on the underlying cognitive processes and knowledge,'' 2000.

\end{thebibliography}
\end{document}